\documentclass[12pt]{article}
\usepackage{epsfig}
\usepackage{graphics}
\newcommand{\comment}[1]{}
\def\be{\begin{equation}}
\def\ee{\end{equation}}
\def\beq{\begin{equation}}
\def\eeq{\end{equation}}
\def\bea{\begin{eqnarray}}
\def\eea{\end{eqnarray}}

\def\gv{g_V}
\def\ga{g_A}
\def\rev1{{\rm Re}\ V_1}
\def\rea1{{\rm Re}\ A_1}
\def\imv1{{\rm Im}\ V_1}
\def\ima1{{\rm Im}\ A_1}
\def\rev2{{\rm Re}\ V_2}
\def\rea2{{\rm Re}\ A_2}
\def\imv2{{\rm Im}\ V_2}
\def\ima2{{\rm Im}\ A_2}
\def\rev3{{\rm Re}\ V_3}
\def\rea3{{\rm Re}\ A_3}
\def\imv3{{\rm Im}\ V_3}
\def\ima3{{\rm Im}\ A_3}
\def\pl{P_L}
\def\plbar{\overline P_L}
\def\pt{P_T}
\def\ptbar{\overline P_T}
\def\peff{P^{\rm eff}_L}

\begin{document}
\begin{flushright}
hep-ph/0605298
\end{flushright}

\begin{center}
\boldmath
{\Large \bf Probing CP-violating contact interactions \\ 
in $e^+e^- \to HZ$ with
polarized beams}
\vskip 1cm
{Kumar Rao\footnote{Email address: kumar@prl.res.in} and Saurabh D.
Rindani\footnote{Email address: saurabh@prl.res.in}}\\
\smallskip
{\it Theory Group, Physical Research Laboratory \\
Navrangpura, Ahmedabad 380009, India}
\vskip 2cm
{\bf Abstract}
\end{center}

\begin{quote}
We examine very general four-point interactions arising due to new
physics contributing to the Higgs production 
process $e^+e^- \to HZ$. We write all 
possible forms for these interactions consistent with Lorentz
invariance. We allow the possibility of CP violation. 
Contributions to the process from anomalous $ZZH$ and
$\gamma ZH$ interactions studied earlier arise as a special
case of our four-point amplitude. Expressions for polar and azimuthal
angular distributions of $Z$ arising from the interference
of the four-point contribution with the standard-model contribution in
the presence of longitudinal and transverse beam polarization are
obtained. An
interesting CP-odd and T-odd contribution is found to be present only
when both electron and positron beams are transversely polarized. Such a
contribution is absent when only anomalous $ZZH$ and
$\gamma ZH$ interactions are considered. We show how angular asymmetries
can be used to constrain CP-odd interactions at a linear collider
operating at a centre-of-mass energy of 500 GeV with transverse beam
polarization.
\end{quote}

\vskip 2cm

\section{Introduction}

Despite the dramatic success of the standard model (SM), an essential component 
of SM responsible for generating masses in the theory, viz., the Higgs 
mechanism, as yet remains untested. The SM Higgs boson, signalling
symmetry breaking in SM by means of one scalar doublet of $SU(2)$, is
yet to be discovered. A scalar boson with the properties of the SM Higgs
boson is likely to be discovered at the Large Hadron Collider
(LHC). However, there are a number of scenarios beyond the standard
model for spontaneous symmetry
breaking, and ascertaining the mass and other properties of the
scalar boson or bosons is an important task. This task would prove
extremely difficult for LHC. However, scenarios beyond SM, with 
more than just one Higgs doublet, as in the case of minimal supersymmetric
standard model (MSSM), would be more amenable to discovery at a linear $e^+e^-$
collider operating at a centre-of-mass (cm) energy of 500 GeV. We are at
a stage when such a linear collider, currently called the International
Linear Collider (ILC), seems poised to become a reality \cite{LC_SOU}. 

Scenarios going beyond the SM mechanism of symmetry breaking, and
incorporating new mechanisms of CP violation have also become a
necessity in order to understand baryogenesis which resulted in the
present-day baryon-antibaryon asymmetry in the universe. In a theory
with an extended Higgs sector and new mechanisms of CP violation, the
physical Higgs bosons are not necessarily eigenstates of CP
\cite{Lee,Pilaftsis}. 
In such a case, the production of a physical Higgs can proceed through
more than one channel, and the interference between two channels can
give rise to a CP-violating signal in the production.

\begin{figure}[htb]
\begin{center}
\epsfig{height=5cm,file=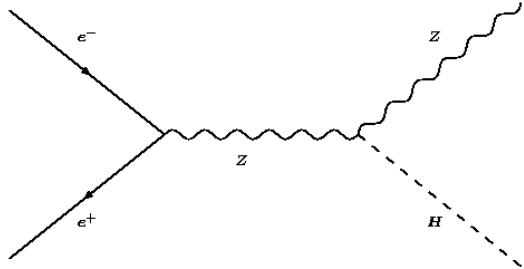}
\end{center}
\caption{Higgs production diagram with an $s$-channel exchange of $Z$
with point-like $ZZH$ coupling.}
\label{fig:vvhptgraph}
\end{figure}
\begin{figure}[htb]
\begin{center}
\epsfig{height=4cm,file=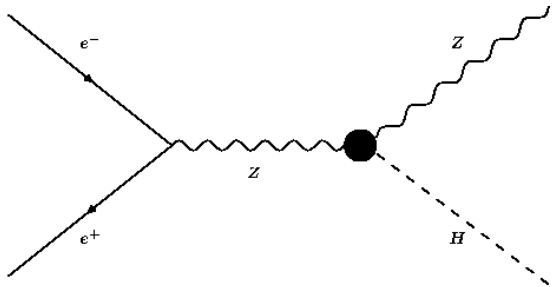}
\end{center}
\caption{Higgs production diagram with an $s$-channel exchange of $Z$
with anomalous $ZZH$ coupling.}
\label{fig:vvhgraph}
\end{figure}
\begin{figure}[htb]
\begin{center}
\epsfig{height=4cm,file=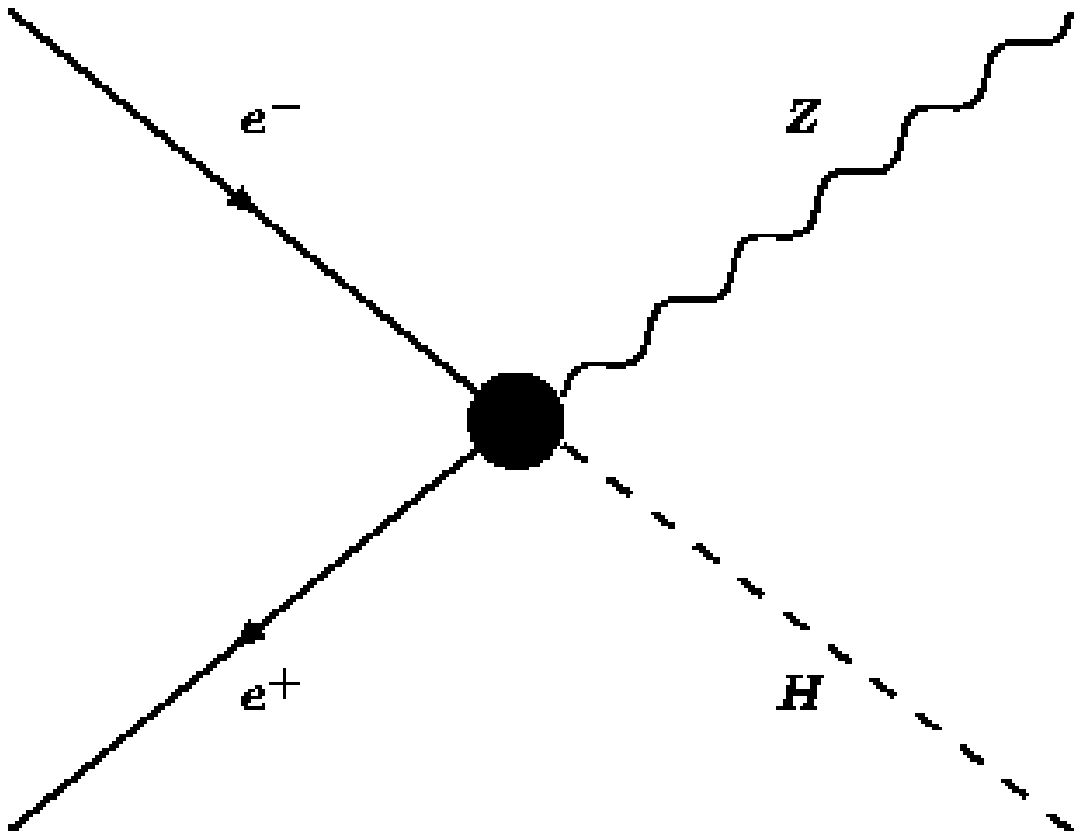}
\end{center}
\caption{Higgs production diagram with a four-point coupling.} 
\label{fig:eehzgraph}
\end{figure}

Here we consider in a general model-independent way the production of a
Higgs mass eigenstate $H$ through the process $e^+e^- \to HZ$. This is
an important mechanism for the production of the Higgs, the other
important mechanisms being $e^+e^- \to e^+e^- H$ and $e^+e^- \to \nu
\overline \nu H$ proceeding via vector-boson fusion. $e^+e^- \to HZ$ is
generally assumed to get a contribution from a diagram with an 
$s$-channel exchange of $Z$. At the lowest
order, the $ZZH$ vertex in this diagram would be simply a point-like
coupling (Fig. \ref{fig:vvhptgraph}).  Interactions
beyond SM can modify this point-like vertex by means of a
momentum-dependent form factor,
as well as by adding more complicated momentum-dependent 
forms of anomalous interactions considered
in \cite{cao,biswal,han,zerwas,gounaris,skjold,hagiwara}. The corresponding diagram
is shown in Fig. \ref{fig:vvhgraph}, where the anomalous $ZZH$ vertex is
denoted by a blob. There could also be a diagram with a photon
propagator and an anomalous $\gamma ZH$ vertex, which we do not show
separately.
We consider here a beyond-SM
contribution represented by a four-point coupling shown in Fig.
\ref{fig:eehzgraph}. This is general enough to include the effects of
the diagram in Fig. \ref{fig:vvhgraph}. Such a discussion would be
relevant in studying effects of box diagrams with new particles, 
or diagrams with $t$-channel exchange of new particles, in addition to
$s$-channel diagrams.

We write down the most general form for the four-point 
coupling consistent with Lorentz invariance. We do not assume CP
conservation.
We then obtain angular distributions for $Z$ (and therefore 
for $H$) arising from the square of amplitude $M_1$ for the diagram in Fig.
\ref{fig:vvhptgraph} with a point-like $ZZH$ coupling, 
together with  the cross term between $M_1$ and the
amplitude $M_2$ for the diagram Fig. \ref{fig:eehzgraph}. We neglect the square of
$M_2$, assuming that this new physics contribution is small compared to
the dominant contribution $\vert M_1 \vert^2$.
We include the possibility that the beams have polarization, either
longitudinal or transverse.
While we have restricted the actual calculation to SM couplings in
calculating $M_1$, it should be borne in mind that 
in models with more than one Higgs doublet this
amplitude would differ by an overall factor depending on the mixing
among the Higgs doublets. Thus our results are trivially applicable to
such extensions of SM, by an appropriate rescaling of the coupling.

We are thus addressing the question of how well the form factors for the
four-point $e^+e^-HZ$ coupling can be determined from the observation of
$Z$ angular distributions in the presence of unpolarized beams or beams
with either longitudinal or transverse polarizations. A similar question
taking into account a new-physics contribution which merely modifies the
form of the 
$ZZH$ vertex has been addressed before in several works
\cite{cao,biswal,han,zerwas,gounaris,skjold,hagiwara}. Those 
works which do take into account four-point couplings, do not do so in all 
generality, but
stop at the lowest-dimension operators \cite{zerwas}. Studies which
include beam polarization in the context of a general $VVH$ vertex 
are \cite{cao,gounaris,hagiwara}.
The approach we adopt here has been used for the process $e^+e^- \to
\gamma Z$ in \cite{basdr,AL1} and for the process $Z\to b \overline b
\gamma$ in \cite{AL2}. A more general analysis of a one-particle
inclusive final state is carried out in \cite{dr}.

The four-point couplings, in the limit of vanishing electron mass, can
be neatly divided into two types -- chirality-conserving (CC) ones and
chirality-violating (CV) ones. The CC couplings involve an odd number of
Dirac $\gamma$ matrices sandwiched between the electron and positron spinors,
whereas the CV ones come from an even number of Dirac $\gamma$ matrices.
In this work, we obtain angular distributions for both CC and CV
couplings. However, since in practice, CV couplings are usually
proportional to the fermionic mass (in this case the electron mass),  
we concentrate on the CC ones (see, however, \cite{lepto}). 

Polarized beams are likely to be available at a linear collider, and
several studies have shown the importance of linear
polarization in reducing backgrounds and improving the sensitivity to
new effects \cite{gudi}. The question of whether transverse beam
polarization, which could be obtained with the use of spin rotators,
would be useful in probing new physics, has been addressed in recent
times in the context of the ILC 
\cite{basdr,lepto,gudi,rizzo,basdrtt,basdrzzg,bartl}.
In earlier work, it has been observed that polarization does not give
any new information about the anomalous $ZZH$ couplings when they are
assumed real \cite{hagiwara}. 
However, in our
work, we find that there are terms in the differential cross section
which are absent unless both electron and positron beams are transversely
polarized. Thus, transverse polarization, if available at ILC, would be
most useful in isolating such terms. This is particularly significant
because these terms are CP violating. Moreover, one of them is even under 
naive
CPT, and thus would survive even when no imaginary part is present in
the amplitude. We discuss the ramifications of this in due course. 

In the next section we write down the possible model-independent
four-point $e^+e^-HZ$ couplings. In Section 3, we obtain the angular
distributions arising from these couplings in the presence of beam
polarization. Section 4 deals with angular asymmetries which can be used
for separating various form factors and Section 5 describes the numerical
results.  Section 6
contains our conclusions and a discussion.

\section{\boldmath Form factors for the process $e^+e^- \to HZ$}

The most general four-point vertex for the process 
\begin{equation}\label{process}
e^- (p_1) + e^+ (p_2) \to Z^\alpha (q) + H(k)
\end{equation}
consistent with Lorentz invariance can be written as
\begin{equation}\label{vertices}
\Gamma^\alpha_{\rm 4pt} =  \Gamma^\alpha_{\rm CC} + 
				\Gamma^\alpha_{\rm CV},
\end{equation}
where the chirality-conserving part $\Gamma^\alpha_{\rm CC}$
containing an odd number of Dirac $\gamma$ matrices is
\def \slZ{Z{\hskip -1.5ex}/}
\def \slq{q{\hskip -1.2ex}/}
\def \slpart{\partial{\hskip -1.25ex}/}
\def \dsp{\displaystyle}

\comment{ This is the old equation:
\begin{equation}\label{vertexcc}
\Gamma^\alpha_{\rm CC} = \frac{i}{M} \gamma^\alpha (V_1 + \gamma_5 A_1)
			-\frac{i}{M^3}\slq (V_2 + \gamma_5 A_2) k^\alpha
			-\frac{1}{M^3}\slq (V_3 + \gamma_5 A_3) (p_2 - p_1)^ 
				\alpha ,
\end{equation}
}
\begin{equation}\label{vertexcc}
\Gamma^\alpha_{\rm CC} = -\frac{1}{M} \gamma^\alpha (V_1 + \gamma_5 A_1)
			+\frac{1}{M^3}\slq (V_2 + \gamma_5 A_2) k^\alpha
			-\frac{i}{M^3}\slq (V_3 + \gamma_5 A_3) (p_2 - p_1)^ 
				\alpha ,
\end{equation}
and the chirality violating part containing an even number of Dirac
$\gamma$ matrices is
\comment{This is the old equation
\begin{eqnarray}\label{vertexcv}
\Gamma^\alpha_{\rm CV}& =&\dsp \frac{1}{M^2}\left[- (S_1 + i \gamma_5 P_1)
				k^\alpha + i (S_2 + i \gamma_5 P_2) 
				(p_2 - p_1)^\alpha \right]\nonumber\\
			&&\!\!\!\! +\dsp
				\frac{i}{M^4}
			\epsilon^{\mu\nu\alpha\beta}p_{2\mu}p_{1\nu}k_\beta 
			(S_3 + i \gamma_5 P_3) .
\end{eqnarray}
}
\begin{eqnarray}\label{vertexcv}
\Gamma^\alpha_{\rm CV}& =&\dsp \frac{i}{M^2}\left[- (S_1 + i \gamma_5 P_1)
				k^\alpha -  (S_2 + i \gamma_5 P_2) 
				(p_2 - p_1)^\alpha \right]\nonumber\\
			&&\!\!\!\! -\dsp
				\frac{1}{M^4}
			\epsilon^{\mu\nu\alpha\beta}p_{2\mu}p_{1\nu}k_\beta 
			(S_3 + i \gamma_5 P_3) .
\end{eqnarray}
In the above expressions, $V_i$, $A_i$, $S_i$ and $P_i$ are form
factors, and are Lorentz-scalar functions of the Mandelstam variables
$s$ and $t$ for the process eq. (\ref{process}). For simplicity, we will only 
consider the case here when the form factors are constants. $M$ is a
parameter with dimensions of mass, put in to render the form factors
dimensionless.

The expressions for the four-point vertices may be thought to arise from
effective Lagrangians
\begin{eqnarray}\label{Lcc}
{\cal L}_{\rm CC}& =& 
			\frac{1}{M} \bar \psi \slZ (v_1 + \gamma_5 a_1) 
                        \psi \phi \nonumber\\
	&&\!\!\!\!		+
			\frac{1}{M^3} \bar \psi \slpart Z^\alpha 
			(v_2 + \gamma_5 a_2) 
                        \psi \partial_\alpha \phi \nonumber \\
	&&\!\!\!\!	+  \frac{i}{M^3} \left[\partial_\alpha \bar \psi
                        \gamma^\mu (v_3 + \gamma_5 a_3) \psi 
			- \bar \psi \gamma^\mu (v_3 + \gamma_5 a_3)
			  \partial_\alpha\psi \right]
                         \phi \partial_\mu Z^\alpha ,
\end{eqnarray}
and
\begin{eqnarray}\label{Lcv}
{\cal L}_{\rm CV}& =&  \frac{1}{M^2} \bar \psi  (s_1 + i\gamma_5 p_1) 
                        \psi \partial_\alpha\phi Z^\alpha\nonumber\\
	&&\!\!	\!\!	+  \frac{i}{M^2} \left[\partial_\alpha \bar \psi
                         (s_2 +i \gamma_5 p_2) \psi 
			- \bar \psi  (s_2 + i \gamma_5 p_2)
			  \partial_\alpha\psi \right] \phi
				Z^\alpha \nonumber\\
 	&&\!\!\!\!		+ 
			\frac{i}{M^4} \epsilon^{\mu\nu\alpha\beta} \partial_\mu 
			\bar \psi (s_3 + i \gamma_5 p_3) \partial_\nu \psi 
			\partial_\beta \phi Z_\alpha  ,
\end{eqnarray}
where the coupling constants $v_i$, $a_i$, $s_i$ and $p_i$ in the
Lagrangians have been promoted to form factors in momentum space when
writing the vertex functions $\Gamma$.

It may be appropriate to contrast our approach with the usual effective
Lagrangian approach. In the latter approach, it is
assumed that SM is an effective theory which is valid up to a cut-off scale
$\Lambda$. The new physics occurring above the scale of the cut-off may
be parametrized by higher-dimensional operators, appearing with powers
of $\Lambda$ in the denominator. These when added to the SM Lagrangian
give an effective low-energy Lagrangian
where, depending on the scale of the momenta involved, one
includes a range of higher-dimensional operators up to a certain maximum
dimension. Our effective theory is not a low-energy limit, so that the
form factors we use are functions of momentum not restricted to low
powers. Thus, the $M$ we introduce is not a cut-off scale, but an
arbitrary parameter, introduced just to make the form factors
dimensionless. 

We thus find that there are 6 independent form factors in the chirality
conserving case, and 6 in the chirality
violating case. An alternative form for the $\Gamma$ above would be
using Levi-Civita $\epsilon$ tensors whenever a $\gamma_5$ occurs. The
independent form factors then are then some linear combinations of the
form factors given above. However, the total number of independent form
factors remains the same.

Note that we have not imposed CP conservation in the above. 
The CP properties of the various terms appearing in the four-point
vertices may be deduced from the CP properties of the corresponding
terms in the effective Lagrangian. Thus, one can check that the terms
corresponding to the couplings $v_3$, $a_3$, $s_1$, $p_2$ and $s_3$ in
the effective Lagrangian are CP violating. As a consequence, the 
terms corresponding to 
$V_3$, $A_3$, $S_1$, $P_2$ and $S_3$ are CP violating, whereas the
remaining are CP conserving. This conclusion assumes that the form factors are 
constants, since the couplings in the effective Lagrangian are
constants. The conclusion can also be carried over when the form factors
are arbitrary functions of $s$ and even
functions of $t-u\equiv \sqrt{s}\vert \vec q \vert \cos\theta$, where
$\theta$ is the angle between $\vec q$ and $\vec p_1$ (or constants). 
This is because in momentum space, $s\equiv (p_1+p_2)^2$ is even under CP,
whereas $t-u\equiv \sqrt{s}\vert \vec q \vert \cos\theta$ is odd under
C and even under P, and thus odd under CP.

The expression for the amplitude for (\ref{process}), arising from the SM 
diagram of Fig. \ref{fig:vvhptgraph} with a point-like $ZZH$ vertex, is 
\begin{equation}\label{smamp}
M_{\rm SM} =- \frac{e^2}{4\sin^2\theta_W \cos^2\theta_W}
\frac{m_Z}{s-m_Z^2} \overline v (p_2) \gamma^\alpha (\gv - \gamma_5 \ga)
u(p_1),
\end{equation}
where the vector and axial-vector couplings of the $Z$ to electrons are
given by
\beq\label{gvga}
\gv= -1 + 4 \sin^2\theta_W,\; \ga=-1,
\eeq
and $\theta_W$ is the weak mixing angle. 
This corresponds to the special case with the following form factors
nonzero:
\begin{equation}\label{smv1}
V_1 = \frac{e^2}{4\sin^2\theta_W \cos^2\theta_W}\frac{Mm_Z}{s-m_Z^2}\gv,
\end{equation}
and 
\begin{equation}\label{sma1}
A_1 = -\frac{e^2}{4\sin^2\theta_W \cos^2\theta_W}\frac{Mm_Z}{s-m_Z^2}\ga.
\end{equation} 
As mentioned earlier, in other models with extra scalar doublets, the
above expressions would be modified simply by a factor depending on the
mixing among the doublets.

\section{Angular distributions}

Using the expression (\ref{smamp}) for the SM contribution from the
diagram (\ref{fig:vvhptgraph}), we now calculate the angular distribution
arising from the square of the SM amplitude and from the interference
between the SM amplitude and the amplitude arising from the four-point
couplings of (\ref{vertexcc}) or (\ref{vertexcv}). We ignore terms
bilinear in the four-point couplings, assuming that the new-physics
contribution is small. We treat the two cases of longitudinal and
transverse polarizations for the electron and positron beams separately.

We choose the $z$ axis to be the direction of the $e^-$ momentum, and
the $xz$ plane to coincide with the production plane. The positive $x$
axis is chosen, in the case of tranvserse polarization, to be along the
direction of the $e^-$ polarization. We then define $\theta$ and $\phi$
to be te polar and azimuthal angles of the momentum $\vec q$ of the $Z$.

We obtain, for the differential cross section with longitudinal
polarization, the expression
\begin{equation}\label{longcs}
\frac{d\sigma_{\rm L}}{d\Omega} =
\frac{d\sigma_{\rm L}^{\rm SM}}{d\Omega}
		+ \frac{d\sigma_{\rm L}^{\rm CC}}{d\Omega},
\end{equation}
where 
\begin{equation}\label{longsm}
\frac{d\sigma_{\rm L}^{\rm SM}}{d\Omega}
=\frac{\lambda^{1/2}}{64\pi^2s}(1-\pl\plbar)
F^2 \left[ \gv^2 + \ga^2 - 2 \gv\ga \peff \right]
 \left[1 + \frac{\vert\vec q \vert^2 \sin^2\theta}{2m_Z^2}
\right]
\end{equation}
 is the SM contribution, and
\bea\label{longcc}
\frac{d\sigma_{\rm L}^{\rm CC}}{d\Omega}
&=&\frac{\lambda^{1/2}}{64\pi^2s}
\frac{2F}{M}(1-\pl\plbar)\nonumber \\
&\times& \left[
\left\{ (\gv - \peff \ga){\rm Re}V_1 + (\peff \gv -
\ga) {\rm Re} A_1 \right\}
\left(1 + \frac{\vert\vec q \vert^2 \sin^2\theta}{2m_Z^2}
\right)\right. \nonumber\\
&-&\left.  \frac{\sqrt{s}q_0}{M^2}
\left\{\left[ (\gv - \peff \ga){\rm Re}V_2 + (\peff \gv -
\ga) {\rm Re} A_2 \right]\right.
\right.\nonumber \\
&& \left. \left. +
 \left[ (\gv - \peff \ga){\rm Im}V_3 + (\peff \gv -
\ga) {\rm Im} A_3 \right] 
\beta_q \cos\theta \right\}\right.\nonumber \\
&&\left. \times
 \frac{\vert \vec q \vert^2}{2m_Z^2}\sin^2\theta
\right]
\eea
is the contribution of the chirality-conserving couplings. There is no
contribution from the chirality-violating couplings for unpolarized
or longitudinally polarized beams.
In the above, we have used 
\beq\label{F}
F = \frac{m_Z}{s-m_Z^2}\left( \frac{e}{2 \sin\theta_W \cos\theta_W}
\right)^2,
\eeq
\beq
\lambda = 4\vert \vec q\vert^2 s = (s-m_H^2-m_Z^2)^2 - 4 m_H^2 m_Z^2,
\eeq
\be
\beta_q = \frac{ \vert \vec q \vert} { q_0},
\ee
and
\be
\peff = \frac{\pl - \plbar}{1 - \pl\plbar}.
\ee

For the case of transverse polarization, we assume that the spins of the
electron and positron are both perpendicular to the beam direction, and
also that they are parallel (or anti-parallel) to each other.
When the beams are transversely polarized we obtain the differential
cross section as

\begin{equation}\label{trancs}
\frac{d\sigma_{\rm T}}{d\Omega} =
\frac{d\sigma_{\rm T}^{\rm SM}}{d\Omega}
		+ \frac{d\sigma_{\rm T}^{\rm CC}}{d\Omega}
		+ \frac{d\sigma_{\rm T}^{\rm CV}}{d\Omega},
\end{equation}
where 
\bea\label{csttran}
\frac{d\sigma_{\rm T}^{\rm SM}}{d\Omega}
&=&\frac{\lambda^{1/2}}{64\pi^2s}
F^2 \left[ (\gv^2 + \ga^2)\left(1 + \frac{\vert\vec q \vert
^2}{2m_Z^2}
\sin^2\theta \right) \right. \nonumber \\ 
&+&\left. \!\! \pt\ptbar (g_V^2-g_A^2) \frac{\vert\vec q
\vert^2}{2 m_Z^2} \sin^2\theta \cos 2 \phi \right]
 \eea
 is the SM contribution,
\beq\label{trancc}
\begin{array}{lcl}
\dsp
\frac{d\sigma_{\rm T}^{\rm CC}}{d\Omega}
&=&\dsp\frac{\lambda^{1/2}}{64\pi^2s}
\frac{2F}{M}
\left\{ 
\left[
(g_V {\rm Re}V_1 - g_A {\rm Re}A_1 )  \right.\right.
\nonumber \\
&+&\dsp
\left. \left.
 \frac{\vert\vec q \vert
^2}{2m_Z^2} \sin^2\theta \left[(g_V {\rm Re}V_1 - g_A {\rm Re}A_1 )
\right. \right. \right. \nonumber \\
&+&\!\!\!\!\dsp\left. \left. \left.  \!\!\pt\ptbar\! 
\left[ (g_V {\rm Re}V_1 + g_A {\rm Re}A_1 ) \cos 2\phi 
- (g_V {\rm Im}A_1 + g_A {\rm Im}V_1 ) \sin 2\phi \right]\right]\right]
\right. \nonumber \\
&-& \dsp\left.\frac{s^{1/2}q_0}{M^2}\frac{\vert \vec q \vert^2}{2m_Z^2}
 \sin^2\theta \left[ (g_V {\rm Re}V_2 - g_A {\rm Re}A_2
) \right.\right. \nonumber \\
&+&\!\!\!\! \left.\left. \!\!\pt\ptbar\! \left[ (g_V {\rm Re}V_2 + g_A {\rm Re}A_2) \cos 2\phi
- (g_V {\rm Im}A_2 + g_A {\rm Im}V_2 ) \sin 2\phi \right]\right]
\right.\nonumber \\
&-& \dsp\left.\frac{s^{1/2}\vert\vec q\vert }{M^2}
\frac{\vert \vec q \vert^2}{2m_Z^2}
\cos\theta \sin^2\theta \left[ (g_V {\rm Im}V_3 - g_A {\rm Im}A_3
) \right.\right. \nonumber \\
&+&\!\!\!\!\left.\left.  \!\!\pt\ptbar\! \left[(g_V {\rm Im}V_3 + g_A {\rm Im}A_3) \cos 2\phi
+ (g_V {\rm Re}A_3 + g_A {\rm Re}V_3 ) \sin 2\phi\right] \right]
\right\}
\end{array}
\eeq
is the contribution from the chirality-conserving couplings, and
\bea\label{trancv}
\frac{d\sigma_{\rm T}^{\rm CV}}{d\Omega}
&=&\frac{\lambda^{1/2}}{64\pi^2s}\frac{Fs^{1/2}}{M^4}\vert \vec q \vert
\sin \theta \nonumber\\
&\times& \left(q_0
\left[-\left\{g_V {\rm Re}
S_1 \sin \phi + g_A {\rm Im} S_1 \cos\phi\right\} (P_T - \ptbar)
 \right. \right. \nonumber \\
&+& \left. \left. \left\{-g_V {\rm Re} P_1 \cos \phi + g_A {\rm Im} P_1
\sin \phi \right\} (P_T + \ptbar) \right]\right.  \nonumber \\
&+&\left. \vert \vec q \vert \cos\theta \left[
\left\{g_A {\rm Re}
S_2 \cos \phi - g_V {\rm Im} S_2 \sin\phi\right\} (P_T - \ptbar)
 \right.\right. \nonumber \\
&-& \left. \left. \left\{g_A {\rm Re} P_2 \sin \phi + g_V {\rm Im} P_2
\cos \phi \right\} (P_T + \ptbar) \right] \right. \nonumber \\
&+&\left.  \frac{1}{2}s^{1/2}\left[\left\{g_A {\rm Re}
S_3 \sin \phi + g_V {\rm Im} S_3 \cos\phi\right\} (P_T - \ptbar)
 \right. \right. \nonumber \\
&+& \left.\left. \left\{g_A {\rm Re} P_3 \cos \phi - g_V {\rm Im} P_3
\sin \phi \right\} (P_T + \ptbar) \right] \right).
\eea
is the contribution from the chirality-violating couplings.

We now examine how the angular distributions in the presence of
polarizations may be used to determine the various form factors.

\section{Polarization and Angular asymmetries} 

The parametrizations we use for the new-physics interactions have 6
complex couplings (form factors) in the CC and case, and 6 in the CV
case. Thus, there are 12 real parameters to be determined in each case. 
We start by making a simplifying assumption that the form factors we
have written down are only functions of $s$ and $t-u$ (or equivalently
$\cos\theta$). In that case, using the unpolarized distributions, which
have approximately the same form as the SM distribution, viz., $A + B
\sin^2\theta$, except for the $V_3$ and $A_3$ terms, which have a
$\sin^2\theta \cos\theta$ dependence, it is not possible to determine
separately all the terms. The terms proportional to $\sin^2\theta
\cos\theta$ can be determined using a simple forward-backward asymmetry:
\beq\label{fbasy}
A_{\rm FB}(\theta_0) =  \frac{1}{\sigma(\theta_0)} 
\left[
\int^{\pi/2}_{\theta_0} \frac{d\sigma}{d\theta}d\theta
-\int^{\pi-\theta_0}_{\pi/2} \frac{d\sigma}{d\theta}d\theta
\right],
\eeq
where 
\beq
\sigma(\theta_0) = \int^{\pi-\theta_0}_{\theta_0}
\frac{d\sigma}{d\theta}d\theta,
\eeq
and $\theta_0$ is a cut-off in the forward and backward directions
needed to keep away from the beam pipe, which could nevertheless be
chosen to optimize the sensitivity.
This asymmetry is odd under CP and is 
proportional to the combination $ \gv {\rm Im}V_3  - \ga
{\rm Im} A_3$. An observation of $A_{\rm FB}(\theta_0)$ can thus
determine that combination of parameters. It should be noted that only
imaginary parts of $V_3$ and $A_3$ enter. This can be related to the
fact that the CP-violating asymmetry $A_{\rm FB}(\theta_0)$ is odd
under naive CPT. It follows that for it to have a non-zero value, the
amplitude should have an absorptive part.  

We now treat the cases of longitudinally and transversely polarized
beams.

\vskip .2cm
\noindent {\it Case(a) Longitudinal polarization:}

The forward-backward asymmetry of eq. (\ref{fbasy}) in the presence of
longitudinal polarization,  which we denote by $A_{\rm
FB}^{\rm L}(\theta_0)$, determines a different
combination of the same couplings ${\rm Im}V_3$ and 
${\rm Im} A_3$. Thus observing asymmetries with and
without polarization, the two imaginary parts can be determined
independently.

In the same way, a combination of the cross section for the unpolarized
and longitudinally polarized beams can be used to determine two
different combinations of the
remaining couplings which appear in (\ref{longcc}). However, one can get
information only on the real parts of $V_1,A_1,V_2$ and $A_2$, not on
their imaginary parts.

With unpolarized or longitudinally polarized beams, it is not possible
to get any information of the chirality-violating couplings, as they do
not contribute.

\vskip .2cm
\noindent{\it Case(b) Transverse polarization:}
\nopagebreak

In the case of the angular distribution with transversely polarized
beams, there is a dependence on the azimuthal angle $\phi$ of the $Z$.
Thus, in addition to $\phi$-independent terms which are the same as
those in the unpolarized case, there are terms with factors $\sin^2\theta
\cos 2 \phi$, $\sin^2\theta \sin  2 \phi$, $\sin^2\theta \cos\theta\cos
2 \phi$ and $\sin^2\theta \cos\theta \sin 2 \phi$ in the case of CC
couplings, and factors $\sin\theta
\cos  \phi$,  $\sin\theta \sin   \phi$,$\sin\theta\cos \theta  \cos \phi$,  
$\sin\theta \cos \theta \sin \phi$
in the case of CV couplings. The $\phi$-dependent terms in the CC case
occur with the factor of $\pt\ptbar$ and in the CV case with a factor of
$\pt + \ptbar$ or $\pt - \ptbar$. Thus, in the CC case, both beams need
to have transverse polarization for a nontrivial azimuthal dependence. 
In the CV case, it is possible to have $\phi$ dependence with either the
electron or the positron beam polarized. We find that that with
the possibility of flipping transverse polarization of one beam, 
it is possible to
examine 4 types of angular asymmetries in each of  CC and CV cases.
Each angular asymmetry would enable the determination of a different 
combination of couplings. 

We will concentrate on the CC case, as most theories permit only CC
couplings, at least in the limit of $m_e=0$. 
We further restrict ourselves here only to terms which involve a $\cos\theta$
factor, which gives rise to a forward-backward asymmetry, due to the
fact that $\cos\theta$ changes sign under $\theta \to \pi - \theta$.
These correspond to the case of CP violation. 

We can then define two different asymmetries, which serve to measure
two different combinations of CP-violating couplings:
\beq\label{atfb}
\begin{array}{lcrrl}
A_{\rm FB}^{\rm T}(\theta_0)& =& \dsp\frac{1}{\sigma(\theta_0)}\!\!&\dsp\left[
\int^{\pi/2}_{\theta_0}d\theta \right.
&\left.
\dsp\left(\int^{\pi/2}_{0} d\phi  - \int^{\pi}_{\pi/2} d\phi 
\right. \right.  \\
&&&& \left.\left.\dsp
+ \int^{3\pi/2}_{\pi} d\phi  - \int^{2\pi}_{3\pi/2} d\phi \right)
\dsp\frac{d\sigma}{d\theta d\phi}\right. \\
&&\left.  -\!\!\right.&\left.   
\dsp\int^{\pi-\theta_0}_{\pi/2}d\theta  \right.  
&\left.
\dsp\left(\int^{\pi/2}_{0} d\phi  - \int^{\pi}_{\pi/2} d\phi 
\right. \right.  \\
&&&& \left.\left.\dsp
+ \int^{3\pi/2}_{\pi} d\phi  - \int^{2\pi}_{3\pi/2} d\phi \right)
\dsp\frac{d\sigma}{d\theta d\phi}
\right],
\end{array}
\eeq
and
\beq\label{atfb2}
\begin{array}{lcrrl}
A_{\rm FB}^{'\rm T}(\theta_0)& =& \dsp\frac{1}{\sigma(\theta_0)}\!\!&\dsp\left[
\int^{\pi/2}_{\theta_0}d\theta \right.
&\left.
\dsp\left(\int^{\pi/4}_{-\pi/4} d\phi  - \int^{3\pi/4}_{\pi/4} d\phi 
\right. \right.  \\
&&&& \left.\left.\dsp
+ \int^{5\pi/4}_{3\pi/4} d\phi  - \int^{7\pi/4}_{5\pi/4} d\phi \right)
\dsp\frac{d\sigma}{d\theta d\phi}\right. \\
&&\left.  -\!\!\right.&\left.   
\dsp\int^{\pi-\theta_0}_{\pi/2}d\theta  \right.  
&\left.
\dsp\left(\int^{\pi/4}_{-\pi/4} d\phi  - \int^{3\pi/4}_{\pi/4} d\phi 
\right. \right.  \\
&&&& \left.\left.\dsp
+ \int^{5\pi/4}_{3\pi/4} d\phi  - \int^{7\pi/4}_{5\pi/4} d\phi \right)
\dsp\frac{d\sigma}{d\theta d\phi}
\right],
\end{array}
\eeq
The former is odd under naive time reversal, whereas the latter is even.
The CPT theorem then implies that these would be respectively dependent
on real and imaginary parts of form factors. 
The integrals in the above may be evaluated to yield
\beq\label{atfbexp}
A_{\rm FB}^{\rm T}(\theta_0)= 3 \pt\ptbar \frac{ \vert\vec q\vert^3 s^{1/2} 
(\gv \rea3 + \ga \rev3) \cos\theta_0 \left( \cos(
          2 \theta_0) - 3
\right)}
{F(\ga^2 + \gv^2)) M^3 \pi (12 m_Z^2
+ 
        5\vert \vec q \vert^2 - \vert \vec q \vert^2 \cos(2 \theta_0))},
\eeq
and
\beq\label{atfb2exp}
A_{\rm FB}^{' \rm T}(\theta_0) = 3  \pt
\ptbar \frac{\vert \vec q\vert^3 s^{1/2}(\ga\ima3 +
\gv\imv3)\cos\theta_0(\cos( 2 \theta_0) - 3)}
	{F(\ga^2 + \gv^2) M^3 \pi
	(12 m_Z^2 + 
	        5\vert \vec q \vert^2 - \vert \vec q \vert^2 \cos(2
		\theta_0))}.
\eeq
We see that the two asymmetries $A_{\rm FB}^{\rm T}$ and 
$A_{\rm FB}^{' \rm T}$ can measure, respectively, the combinations $\gv
\rea3 + \ga \rev3$ and $\ga\ima3 + \gv\imv3$. The latter is dominated by
$\ima3$ which may also be determined using unpolarized beams. The former
requires transverse polarization to measure.

It can be checked that if one considers only contributions from a
modification of the $ZZH$ vertex as in
\cite{cao,biswal,han,zerwas,gounaris,skjold,hagiwara}
$A_{\rm FB}^{\rm T}$
and $A_{\rm FB}^{' \rm T}$ vanish. This result for $A_{\rm FB}^{\rm T}$ 
is obtained in \cite{hagiwara}. 
Thus, observation of a
nonzero asymmetry would signal the presence of the CP-violating
four-point interaction.  

\section{Numerical Results}

We now obtain numerical results for the polarized cross sections, the
asymmetries and the sensitivities of these asymmetries for a definite
configuration of the linear collider. For our numerical calculations, we
have made use of the following values of parameters: $m_Z = 91.19$~GeV,  
$\alpha = 1/128$,  
$\sin^2\theta_W = 0.22$, $M = 1$~TeV. It should be noted that the
particular choice of $M$ is simply for convenience, and is 
not simply related to any assumption about the
scale of new physics -- a change in $M$ can always be compensated by
corresponding changes in the form factors. For the parameters of the linear
collider, we have assumed $\sqrt{s}=500$~GeV,
  $\pl = 0.8$,  $\plbar = -0.6$,   $\pt = 0.8$, $\ptbar = 0.6$, and an
  integrated luminosity $\int \mathcal{L} dt = 500\ \textrm{fb}^{-1}$.
  For most of our calculations we choose three values of the Higgs mass,
  $m_H=150$~GeV, 200~GeV and 300~GeV.

We have assumed that the contribution of the $Z$ exchange diagram of
Fig. \ref{fig:vvhptgraph} is the same as that in SM.  Since we are
keeping open the possibility that the Higgs boson we are dealing with is
not an SM Higgs, this assumption may not be correct. However, the
modification for a Higgs of a different model will be multiplication by
a certain overall factor depending on the mixing of the different Higgs
bosons in the model. This can easily be taken care of while interpreting
our results for such a model.

\begin{figure}[htb]
\input{plot1.tex}
\caption{The SM cross section for longitudinally polarized beams with
$P_L=0.8$, $\plbar= -0.6$ integrated over the polar angle in the
range $\theta_0<\theta<\pi - \theta_0$ as a function of $\theta_0$ for
different values of $m_H$.}\label{fig:longcssm} 
\end{figure}
In Fig. \ref{fig:longcssm} we have plotted the SM cross section in the
presence of longitudinally polarized beams as a function of the cut-off
$\theta_0$ for three values of $m_H$.
Fig. \ref{fig:trancssm} shows the corresponding plot for the SM cross
sections with transversely polarized beams.
\begin{figure}[htb]
\input{plot2.tex}
\caption{The SM cross section for transversely polarized beams with
$\pt=0.8$, $\ptbar= 0.6$ integrated over the polar angle in the
range $\theta_0<\theta<\pi - \theta_0$ 
as a function of $\theta_0$ for different values of $m_H$.}\label{fig:trancssm} 
\end{figure}

\begin{figure}[htb]
\vskip .3cm
\input{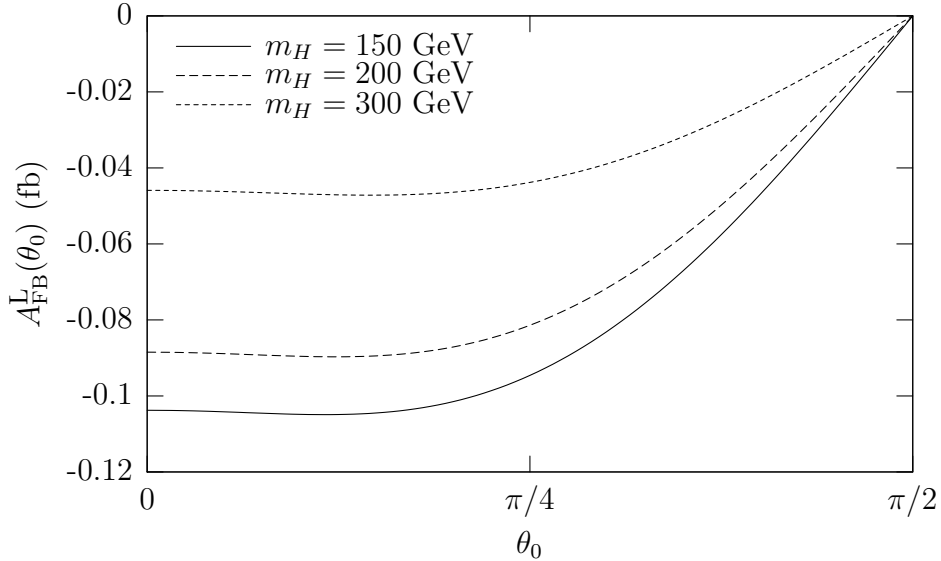}
\caption{The forward-backward asymmetry for longitudinally polarized
beams, $A^{\rm L}_{\rm FB}(\theta_0)$, 
for $\imv3=0.1$ as a function of $\theta_0$. All other couplings are
taken to be zero.}\label{fig:longfb}
\end{figure}

\begin{figure}[htb]
\input{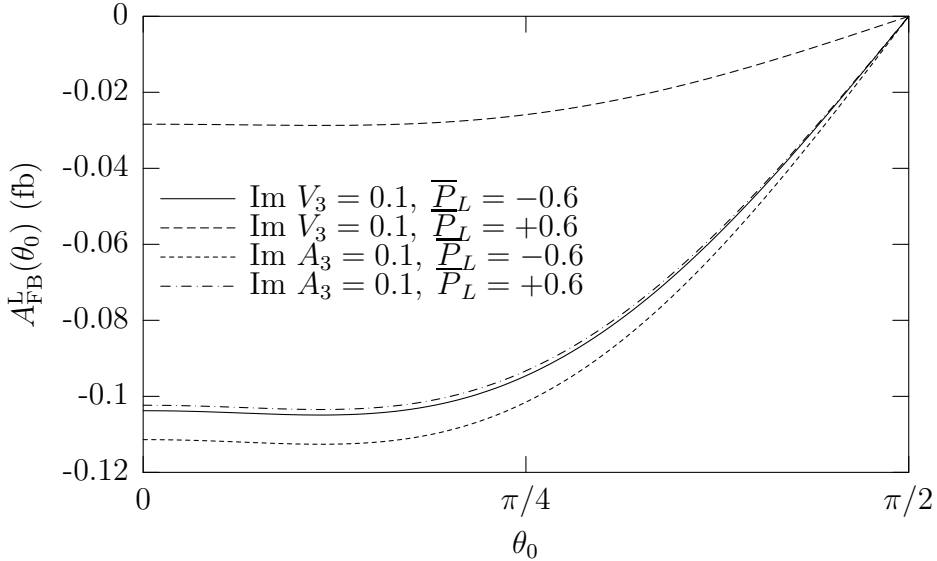}
\caption{The asymmetry $A^{\rm L}_{\rm FB}(\theta_0)$ defined in the
text for $m_H=150$~GeV, $\pl=0.8$, for two combinations of 
couplings, and for two values of $\plbar$ differing in sign. 
}\label{fig:longfb2}
\end{figure}

In Fig. \ref{fig:longfb} is plotted the forward-backward asymmetry
$A^{\rm L}_{\rm FB}(\theta_0)$ with longitudinal polarization as a
function of the cut-off $\theta_0$. Only the  parameter $\imv3$ is chosen 
nonzero, and to have the value 0.1. This choice is for illustration. 
Fig. \ref{fig:longfb2} shows the same asymmetry for a fixed value of
$m_H=150$~GeV, for the combinations $\imv3 = 0.1,\; \ima3=0$ and
$\imv3 = 0,\; \ima3=0.1$, and for values of $\plbar$
differing in sign. 
The asymmetry depends on the relative signs of $\pl$ and $\plbar$ 
and is larger in magnitude when the relative signs are opposite.

In the case of transverse polarization, the two asymmetries $A_{\rm
FB}^{\rm T}$ 
and $A_{\rm FB}^{' \rm T}$ are shown as functions of
$\theta_0$ in Figs.  \ref{fig:tranfb} and \ref{fig:tranfb2}
for values of polarization $\pt =  0.8$ and $\ptbar = 0.6$. 
In Fig.  \ref{fig:tranfb}, the only nonzero
parameter is $\rev3=0.1$, whereas in Fig. \ref{fig:tranfb2},
the only nonzero parameter is $\imv3=0.1$.
\begin{figure}[htb]
\input{plot8.tex}
\caption{The asymmetry $A_{\rm FB}^{\rm T}$ 
for transverse polarizations $\pt =  0.8$ and $\ptbar = 0.6$
plotted against $\theta_0$ for $\rev3=0.1$. All other couplings are
taken to be zero.}\label{fig:tranfb}
\end{figure}
\begin{figure}[htb]
\input{plot9.tex}
\caption{The asymmetry $A_{\rm FB}^{' \rm T}$ 
for transversely polarizations $\pt =  0.8$ and $\ptbar = 0.6$
plotted against $\theta_0$ for $\imv3=0.1$. All other couplings are
taken to be zero.}\label{fig:tranfb2}
\end{figure}

In all the above figures, the dependence on the cut-off $\theta_0$ is
mild for small values of $\theta_0$. Hence the results will not be
sensitive to the choice of $\theta_0$, if it is small.

We now examine the accuracy to which each of the couplings can be
determined for linear collider operating at $\sqrt{s}=500$~GeV and with
an integrated luminosity of 500~fb$^{-1}$. At the 90\% confidence level
(CL), the limit that can be placed on a parameter contributing linearly to a
certain asymmetry $A$ is given by $1.64 /(A_1\sqrt{N_{\rm SM}})$, where
$A_1$ is the asymmetry for unit value of the parameter.   

We first consider the determination of the parameter $\rev3$ from a
measurement of the asymmetry $A_{\rm FB}^{\rm T}$ for a typical value of
$\theta_0=45^{\circ}$. If the asymmetry is not observed, we find that
the limit placed on $\rev3$ is $3.9\times 10^{-2}$ for $m_H=150$~GeV,
$5.1\times 10^{-2}$ for $m_H=200$~GeV, and $1.3\times 10^{-1}$ for
$m_H=300$~GeV. Since the combination which appears in the asymmetry is
$\gv \rea3 + \ga \rev3$, it implies that the corresponding limits on
$\rea3$ will be a factor $\vert \ga/\gv \vert \approx 8.3$ higher. Thus, the
asymmetry is more sensitive to $\rev3$ because of a larger coupling
$\ga$ multiplying it.

The expression for $A_{\rm FB}^{' \rm T}$ is identical to that for
$A_{\rm FB}^{\rm T}$, except that the factor $\gv \rea3 + \ga \rev3$ is
replaced by $\gv \imv3 + \ga \ima3$. Thus, now it will be $\ima3$ which
will have the limits mentioned above for $\rev3$, and $\imv3$ will have
limits which are a factor of about 8.3 larger.

It should be borne in mind that the definition of the couplings (form
factors) are dependent on the value of the scale parameter $M$ which is
chosen. Thus changing the value of $M$ will change the limits on the
form factors.

\section{Conclusions and Discussion}

We have parametrized the amplitude for the process $e^+e^- \to HZ$ using
only Lorentz invariance by means of form factors, treating separately
the chirality-conserving and chirality-violating cases. We then
calculated the differential cross section for the process $e^+e^- \to
HZ$ in terms of these form factors for polarized beams. 
The motivation was to determine the  extent to which 
longitudinal and transverse polarizations
can help in an independent determination of the various form factors.

We found that in the presence of transverse polarization, there is a
CP-odd and T-odd contribution to the angular distribution. The coupling
combinations this term depends on cannot be determined using
longitudinally polarized beams. Moreover, this transverse-polarization
dependent contribution does not arise when only $VVH$ type of couplings
are considered. Hence such a term, if observed, would be a unique signal
of CP-violating four-point interaction.

It should be emphasized that our results and conclusions are dependent on
the assumption that the form factors are independent of $t$ and $u$. In
particular, the CP property of a given term in the distribution would
change if the corresponding form factor is an odd function of
$\cos\theta$. The reason is that $\cos\theta \equiv q\cdot
(p_2-p_1)/(\vert \vec q \vert s^{1/2})$ is odd under CP.

We have discussed limits on the couplings that would be expected from a
definite configuration of the linear collider. 
As for the 
CP-conserving couplings, limits may be obtained even from the existing
LEP data, which has excluded SM Higgs up to mass of about 114 GeV.
However, we have concentrated only
on the limits on the CP-violating couplings. 
It should be borne in mind that the limits on these depend on the choice 
of $M$, the arbitrary parameter of dimension of mass that we introduced. 

Though we have used SM couplings for the leading contribution of Fig.
\ref{fig:vvhptgraph}, as mentioned earlier, the analysis needs only trivial
modification when applied to a model like MSSM or a multi-Higgs-doublet
model, and will be useful in such extensions of SM.
It is likely that such models will give rise to four-point contributions
through box diagrams or loop diagrams with a $t$-channel exchange of
particles. However, to our knowledge, such calculations are not
available for
CP-violating models. The interesting effects we have discussed would
make it useful to carry out such calculations.

We have discussed the angular distribution of the $Z$ in the process
$e^+e^- \to HZ$. Clearly, for the discussion to be of practical use, one
has to include the means of detection of $Z$ and $H$. Thus, it is
important to include decays of $Z$ and $H$ and to see what our analysis
implies for the decay products. In particular, one has to answer the
question as to how leptons or jets from the decay  of the $Z$ can be
used to measure the asymmetries we discuss, and with what efficiency.
To the extent that the sum of the four-momenta of the charged lepton
pair or the jet pair can be a measure of the $Z$ four-momentum, it
should be possible to reconstruct the asymmetries discussed here with
reasonable accuracy.
One should also investigate  the effect experimental cuts would
have on the accuracy of the determination of the couplings. 
One should keep in mind the possibility that radiative corrections can
lead to quantitative changes in the above results (see, for example, 
\cite{comelli}).
While these
practical questions are not addressed in this work, 
we feel that the interesting new features we found would make it
worthwhile to address them in future.

\noindent Acknowledgement: This work was partly supported by the IFCPAR project
no. 3004-2. We thank Rohini Godbole and B. Ananthanaryan 
for discussions and  for comments on the manuscript.

  S.~Weinberg,
      Phys.\ Rev.\ Lett.\  {\bf 37} (1976) 657.

\bibitem{Pilaftsis}
  A.~Pilaftsis,
     ``CP-odd tadpole renormalization of Higgs scalar-pseudoscalar
     mixing,''
         Phys.\ Rev.\ D {\bf 58} (1998) 096010
	   [arXiv:hep-ph/9803297].

	       A.~Pilaftsis,
	          ``Higgs scalar-pseudoscalar mixing in the minimal
		  supersymmetric standard
		     model,''
		         Phys.\ Lett.\ B {\bf 435} (1998) 88
			   [arXiv:hep-ph/9805373];

			       D.~A.~Demir,
			          ``Effects of the supersymmetric phases
				  on the neutral Higgs sector,''
				      Phys.\ Rev.\ D {\bf 60} (1999) 055006
				        [arXiv:hep-ph/9901389];

				  A.~Pilaftsis and C.~E.~M.~Wagner,
				     ``Higgs bosons in the minimal
				     supersymmetric standard model with
				     explicit  CP
				        violation,''
					    Nucl.\ Phys.\ B {\bf 553}
					    (1999) 3
					      [arXiv:hep-ph/9902371].
						9902371;

\bibitem{cao}
Q.~H.~Cao, F.~Larios, G.~Tavares-Velasco and C.~P.~Yuan,
arXiv:hep-ph/0605197.

\bibitem{biswal}
S.~S.~Biswal, R.~M.~Godbole, R.~K.~Singh and D.~Choudhury,
Phys.\ Rev.\ D {\bf 73}, 035001 (2006)
[arXiv:hep-ph/0509070].

\bibitem{han}
T.~Han and J.~Jiang,
Phys.\ Rev.\ D {\bf 63}, 096007 (2001)
[arXiv:hep-ph/0011271].

\bibitem{zerwas}
V.~Barger, T.~Han, P.~Langacker, B.~McElrath and P.~Zerwas,
Phys.\ Rev.\ D {\bf 67} (2003) 115001
[arXiv:hep-ph/0301097];
W.~Kilian, M.~Kramer and P.~M.~Zerwas,
arXiv:hep-ph/9605437.

\bibitem{gounaris}
G.~J.~Gounaris, F.~M.~Renard and N.~D.~Vlachos,
Nucl.\ Phys.\ B {\bf 459}, 51 (1996)
[arXiv:hep-ph/9509316].

\bibitem{skjold}
A.~Skjold and P.~Osland,
Nucl.\ Phys.\ B {\bf 453} (1995) 3
[arXiv:hep-ph/9502283].

\bibitem{hagiwara}
K.~Hagiwara and M.~L.~Stong,
Z.\ Phys.\ C {\bf 62}, 99 (1994)
[arXiv:hep-ph/9309248].

\bibitem{basdr}
B.~Ananthanarayan and S.~D.~Rindani,
Phys.\ Lett.\ B {\bf 606} (2005) 107
[arXiv:hep-ph/0410084];
JHEP {\bf 0510}, 077 (2005)
[arXiv:hep-ph/0507037].

\bibitem{AL1}
K.~J.~Abraham and B.~Lampe,
Phys.\ Lett.\ B {\bf 446} (1999) 163
[arXiv:hep-ph/9810205].

\bibitem{AL2}
K.~J.~Abraham and B.~Lampe,
Phys.\ Lett.\ B {\bf 326} (1994) 175.

\bibitem{dr}
B.~Ananthanarayan and S.~D.~Rindani,
arXiv:hep-ph/0601199, Eur. Phys. J. C (to appear).

\bibitem{lepto}
S.~D.~Rindani,
Phys.\ Lett.\ B {\bf 602}, 97 (2004)
[arXiv:hep-ph/0408083].

\bibitem{gudi}
G.~Moortgat-Pick {\it et al.},
arXiv:hep-ph/0507011.

\bibitem{rizzo}
T.~G.~Rizzo,
JHEP {\bf 0302}, 008 (2003)
[arXiv:hep-ph/0211374];

J.~Fleischer, K.~Kolodziej and F.~Jegerlehner,
Phys.\ Rev.\ D {\bf 49}, 2174 (1994);

M.~Diehl, O.~Nachtmann and F.~Nagel,
Eur.\ Phys.\ J.\ C {\bf 32}, 17 (2003)
[arXiv:hep-ph/0306247];

S.~Y.~Choi, J.~Kalinowski, G.~Moortgat-Pick and P.~M.~Zerwas,
Eur.\ Phys.\ J.\ C {\bf 22}, 563 (2001)
[Addendum-ibid.\ C {\bf 23}, 769 (2002)]
[arXiv:hep-ph/0108117].

\bibitem{basdrtt}
B.~Ananthanarayan and S.~D.~Rindani,
Phys.\ Rev.\ D {\bf 70}, 036005 (2004)
[arXiv:hep-ph/0309260];

\bibitem{Rindani:2004wr}
S.~D.~Rindani,
arXiv:hep-ph/0409014.

\bibitem{basdrzzg}
B.~Ananthanarayan, S.~D.~Rindani, R.~K.~Singh and A.~Bartl,
Phys.\ Lett.\ B {\bf 593}, 95 (2004)
[Erratum-ibid.\ B {\bf 608}, 274 (2005)]
[arXiv:hep-ph/0404106];

J.~Kalinowski,
arXiv:hep-ph/0410137;

P.~Osland and N.~Paver,
arXiv:hep-ph/0507185.

\bibitem{bartl}
A.~Bartl, K.~Hohenwarter-Sodek, T.~Kernreiter and H.~Rud,
Eur.\ Phys.\ J.\ C {\bf 36}, 515 (2004)
[arXiv:hep-ph/0403265];

A.~Bartl, H.~Fraas, S.~Hesselbach, K.~Hohenwarter-Sodek, T.~Kernreiter
and G.~Moortgat-Pick,
JHEP {\bf 0601}, 170 (2006)
[arXiv:hep-ph/0510029];

S.~Y.~Choi, M.~Drees and J.~Song,
arXiv:hep-ph/0602131.

\bibitem{comelli}
P.~Ciafaloni, D.~Comelli and A.~Vergine,
JHEP {\bf 0407} (2004) 039
[arXiv:hep-ph/0311260].


\thebibliography{99}
\bibitem{LC_SOU}
T.~Abe {\it et al.}  [American Linear Collider Working Group],
in {\it Proc. of the APS/DPF/DPB Summer Study on the Future of
Particle Physics
(Snowmass 2001) } ed. N.~Graf
arXiv:hep-ex/0106055;
J.~A.~Aguilar-Saavedra {\it et al.}  [ECFA/DESY LC Physics
Working Group]
arXiv:hep-ph/0106315;
K.~Abe {\it et al.}  [ACFA Linear Collider
Working Group]
arXiv:hep-ph/0109166.

	   \bibitem{Lee}
	     T.~D.~Lee,
	        ``CP Nonconservation And Spontaneous Symmetry
		Breaking,''
		  %
		    Phys.\ Rept.\  {\bf 9} (1974) 143;
\end{document}